\documentclass[plb,preprint,tightenlines,floatfix,preprintnumbers,nofootinbib,nofootinbib,amsmath,english]{revtex4}
  \usepackage{amssymb}
   \usepackage{amsmath}
    \usepackage{epsfig}
     \usepackage{bm}
      \usepackage{pifont}
\newcommand{\vecc}[1]{\mbox{\boldmath $#1$}}

\begin{document}

\title{On a Wilson lines approach to the study of jet quenching}
\author{I.O.~Cherednikov}\footnote{Also at: {\it Bogoliubov Laboratory of Theoretical Physics, JINR, RU-141980 Dubna, Russia}}
\email{igor.cherednikov@uantwerpen.be}
\affiliation{Departement Fysica, Universiteit Antwerpen, B-2020 Antwerpen, Belgium\\}
\author{J.~Lauwers}
\email{jasper.lauwers@uantwerpen.be}
\affiliation{Departement Fysica, Universiteit Antwerpen, B-2020 Antwerpen, Belgium\\}
\author{P.~Taels}
\email{pieter.taels@uantwerpen.be}
\affiliation{Departement Fysica, Universiteit Antwerpen, B-2020 Antwerpen, Belgium\\}
\vspace {10mm}
\date{\today}

\begin{abstract}
We address the geometrical structure of the ``skewed'' correlator of two space-like separated (almost) oppositely directed Wilson lines. Similar objects occur in the analysis of the transverse-momentum broadening probability function, the first moment of which is associated with the jet quenching parameter. We start from the Euclidean space formulation and then transform the result to the Minkowski light-cone geometry, arguing that this procedure is consistent in the leading order
of the perturbative expansion. We discuss as well the issues of the UV, rapidity and IR singularities, and possible use of the proposed approach in lattice simulations.
\end{abstract}

\maketitle


\section{Definitions and setting the hierarchy of scales}

Fast partons, created as a result of heavy-ion collisions, propagate through the nuclear medium and lose their energy due to different mechanisms (collinear pair production, Bremsstrahlung, etc.). This class of phenomena is collectively known as jet quenching. The energy loss shows up in the spectrum of high-$p_{\perp}$ hadrons in the final state, as well as in the kinematical parameters of the jets produced by the parton. In the former, the jet quenching can cause a suppression of the spectrum,  while the latter can exhibit an imbalance of the transverse momentum of high-$p_{\perp}$ back-to-back jets, and an increased angular broadening of the final jets \cite{d'Enterria:2009am}. Jet quenching is observed in the heavy ion collisions at RHIC \cite{RHIC_exp} and at the LHC \cite{LHC_exp}. It is assumed that the above mentioned medium is a dense, deconfined state of quarks and gluons is formed in such collisions, known as the quark-gluon plasma (QGP) \cite{QGP_old}. For a recent review of jet quenching, see, e.g., Ref. \cite{d'Enterria:2009am}, while \cite{Martinez:2013xka} provides an up-to-date report on the quark-gluon plasma.

In the leading approximation, the multiple soft interactions of the hard parton with the medium do not change the parton energy, but affect its momentum component transverse to the initial (collinear) direction. Within a jet, there will be therefore no change in mean momentum, but rather an increase in the transverse momentum spread. This effect is known as transverse momentum broadening, and can be characterized by the so-called jet quenching parameter $\hat{q}$ \cite{hat_q_old}:
\begin{equation}
\hat{q}\equiv L^{-1}{\langle \vecc k_{\perp}^{2}\rangle}
=
L^{-1} \int\! d^2 k_\perp \vecc k_{\perp}^{2} \  P (\mathbf{k}_\perp)  \ ,
\label{eq:hat_q_def}
\end{equation}
where $k_{\perp}$ is the acquired transverse momentum after propagation through the medium for a (longitudinal) distance $L$ and $P (\mathbf{k}_\perp)$ is the probability distribution. Dimension of $\hat q$ is, therefore, $[{\rm mass}^3]$, and $\hat q$ measures the average transverse momentum squared in unit of length.
Within the framework of soft-collinear effective theory (SCET) \cite{Bauer:2000yr}, the jet quenching parameter can be computed in general covariant gauge, yielding a thermal correlator of two transversely separated light-like Wilson lines \cite{hat_q_AdS, hat_q_T_line}. Recently it was demonstrated that, by the addition of transverse Wilson lines at light-cone infinity (these objects were first introduced in order to save gauge invariance in the operator definition of the transverse-momentum dependent PDFs, by taking into account the effects of initial/final state interactions \cite{T_line_origin}, and, subsequently, by considering the renormalization-group properties of the TMD matrix elements \cite{CS_T_line} ), this result can be generalized to the light-cone gauge \cite{hat_q_T_line}.

In this paper we study those properties of the jet quenching parameter which are overlooked in the current literature, namely, the geometrical structure of the
corresponding transverse-distance (or momentum) dependent correlator and the issue of continuation from Euclidean to Minkowski space-time.
{For that reason, we concentrate on the non-thermal medium with zero temperature $T=0$. This setup allows us to study the ``cold background'' of the jet quenching process, in contrast to the medium with finite temperature that is created by the nuclear collisions. We take into account, however, that even in the cold medium the correlation functions may be nonperturbative. We will make use of the general covariant two-gluon correlation function, which can contain dependence on an additional low energy scale $Q_T$, so that
\begin{equation}
\lambda = Q_T / Q \ll 1 \ ,
\label{eq:scales_1}
\end{equation}
where $Q$ is the large scale characterizing the momentum of the primordial jet \cite{hat_q_T_line}. The transverse momentum broadening is, therefore, of the order
\begin{equation}
\vecc p_T-{\rm broadening} \sim \lambda Q = Q_T \ ,
\label{eq:scales_2}
\end{equation}
while the virtuality of the parton is proportional to the length of the light-like Wilson line and is of the order
\begin{equation}
{\rm virtuality} \sim L^{-2} \sim (\lambda Q )^2 = Q_T^2 \ .
\label{eq:scales_3}
\end{equation}
Hence we have two independent low-energy scales, which obey the relation $Q_T L \approx 1$. This assumption will be shown below to play a significant role in the scale dependence of the Wilson lines correlator under consideration. {It is worth mentioning that the hierarchy of scales includes unavoidably other, even ``softer'' energy parameters, e.g., $g Q_T, g^2 Q_T$, where $g$ is a coupling constant. In the present work we explore the region between the two first scales, $[gQ_T, Q_T]$, so that $Q_T$ is an effective UV cutoff, while $gQ_T$ is an IR one. } }

Following the results of Ref. \cite{hat_q_T_line}, we assume that the probability distribution $P (\mathbf{k}_\perp)$ exists and can be {\it defined} as the Fourier transform of the expectation value of two light-like Wilson lines (for the sake of simplicity, we take the fields ${\cal A} = A^a \cdot t^a$ in the fundamental representation of the gauge group, hence the factor $1/N_c$)
\begin{equation}
P (\mathbf{k}_\perp; n^-)
=
\int d^{2}z_{\perp} \
{\rm e}^{i k_{\perp} \cdot  z_{\perp}} \  \tilde P (\vecc z_\perp)
=
\int d^{2}z_{\perp} \ {\rm e}^{i k_{\perp} \cdot  z_{\perp}} \ \langle \star |  \frac{1}{N_c} \mathrm{ Tr}\ \{ {\cal W}_{n^-}^{\dagger}[0, \vecc z_{\perp}] {\cal W}_{n^-} [0, \vecc 0_\perp] \} | \star \rangle \ ,
\label{eq:jetquenchingWilson}
\end{equation}
where $| \star \rangle$ denotes the state accumulating information about the properties of the medium, and explicit dependence on the light-like vector ${n^-}$ is included. The generic light-like Wilson line operator evaluated along the light-like direction $y^- = n^- \ \sigma$ reads
\begin{equation}
{\cal W}_{n^-} [ y^{+}, \vecc y_{\perp}]
= {\cal P}\, \exp{\left[ig \int_{-L/2}^{L/2}\! dy^{-} {\cal A}^{+} (y^{+}, y^{-}, \vecc y_{\perp}) \right] } \ ,
\end{equation}
where the length $L$ sets the longitudinal scale of the medium. Note that path-ordering operator ${\cal P}$ orders colour matrices $t^a$, but not time-dependent fields $A^a$. The latter must be ordered by an additional ordering, time- or anti-time, if we want Eq. (\ref{eq:jetquenchingWilson}) to be consistent with the physical situation. This issue is not, however, our concern in the present work since in the $g^2$-order, the diagram contributing to $\tilde P (\vecc z_\perp)$ contains only the contraction of two space-like separated gluon fields, which obviously commute. Of course, in the NLO calculations one has to take the time- and anti-time- ordering into account.

\section{Calculation in Euclidean space}

Perturbative calculations of the Wilson lines and loops call for a careful treatment of the angular dependence (e.g., cusps, self-intersections, etc.) and the possible divergences of various kinds both in Minkowski and Euclidean space-time (see, e.g., Refs. \cite{WL_angular_origin, WL_angular, WL_Eucl}). On the other hand, nonperturbative analysis of the jet quenching is possible, which can be carried out by making use of other methods---see, e.g., \cite{hat_q_AdS, hat_q_NP}. Here we address the issue of the angular dependence of the generic {\it skewed} correlator of two Wilson lines, defined first in Euclidean space, and then show how to transform to the Minkowskian geometry on the light-cone. We do not specify the way how the expectation values in the medium $\langle \star | {\cal W}^\dag {\cal W} | \star \rangle$ must be evaluated, instead we try to retrieve as much as possible information, making as less as possible conjectures about the properties of the two-gluon contraction in a medium.

To this end, let us consider the following object defined in Euclidean space
\begin{equation}
\tilde P (\vecc z_\perp; v, \bar v)
=
 \langle \star |  \frac{1}{N_c} \mathrm{ Tr}\ \{ {\cal W}^{\dagger}_{\bar v}[\vecc z_{\perp}] {\cal W}_{v} [\vecc 0_\perp] \} | \star \rangle \ ,
\label{eq:P_Euclid_def}
\end{equation}
where
\begin{equation}
{\cal W}_{v} [\vecc 0_\perp]
=
{\cal P}\, \exp{\left[ig v_\mu \ \int_{-\infty}^{\infty}\! d\sigma {\cal A}_\mu (y) \right] } \ , \
y = v \sigma \ ,
\end{equation}
and
\begin{equation}
{\cal W}_{\bar v}^\dag [\vecc z_\perp]
=
{\cal P}\, \exp{\left[- ig  \bar v_\mu  \
\int_{-\infty}^{\infty}\! d\sigma' {\cal A}_\mu (y') \right] } \ , \
y' = \bar v \sigma'  + \vecc z_\perp\ ,
\end{equation}
where the directions of the Euclidean vectors $v$ and $\bar v$ are determined by the angles $\phi, \bar \phi$
\begin{eqnarray}
& & v_\mu = (v^0, v^z, \vecc 0_\perp) = L (\cos \phi/2, \sin \phi/2 , \vecc 0_\perp ) \ , \\
& & \bar v_\mu = (\bar v^0, \bar v^z, \vecc 0_\perp) =
- L (\cos \bar \phi/2, \sin \bar \phi/2 , \vecc 0_\perp ) \\
& & v^2 = \bar v^2 = L^2 \ .
\end{eqnarray}
This object is more general than is needed for the straightforward calculation of the jet quenching parameter.
We define as well an asymmetric function of the Euclidean vectors $(v, \bar v)$
\begin{equation}
 \rho (v, \bar v)
 =
 \frac{1}{L} \ \int\! d^2 k_\perp \ \vecc k_\perp^2 \ \int d^{2}z_{\perp} \ {\rm e}^{i k_{\perp} \cdot  z_{\perp}} \ \tilde P (\vecc z_\perp; v, \bar v) \ ,
\label{eq:rho_def}
\end{equation}
which formally arises as the skewed analogue of the physical $\hat q$, so that we assume that there exists an appropriate transition procedure
\begin{equation}
\rho (v, \bar v)
\to
\hat q_{\rm LC} \ .
\end{equation}
However, it is not our concern in the present work to specify this procedure.

Naively, the realistic situation is supposed to be achieved by making the transformation of the angles to the Minkowski geometry $(\phi, \bar \phi) = i (\psi, \bar \psi)$ and setting them equal. The light-cone case can be obtained, formally, by taking the limit of large Minkowskian angles $\psi$ and $\bar \psi$. We will see, however, that this straightforward strategy does not work in our case. Instead we will keep the two angles different after transformation to Minkowski space-time and, given that the angular dependence gets factorized into a covariant multiplier, demonstrate that the light-cone limit can be consistently performed in the skewed layout.

Another important change as compared to the standard definition of $P (\vecc k_\perp)$ is that we evaluate the line integrals in the Wilson functionals along the {\it infinite}  paths, keeping IR singularities under control, if needed, by an additional energy scale $\Lambda \sim g Q_T$. In the dimensional regularization, $\Lambda$ is introduced formally as an energy parameter in the integration measure. On the other hand, the length $L$ provides the natural longitudinal scale. Recall that the length of an integration contour in the coordinate space corresponds to the inverse virtuality of the eikonalized particle in the momentum space $L \sim m^{-1}$ \cite{WL_angular}.
Let us note that the contribution under consideration is UV finite due to the space-like separation of the Wilson lines. This is not the case anymore in the NLOs.

The leading non-trivial term of the weak-field expansion of the skewed probability distribution (\ref{eq:P_Euclid_def}) reads, see Fig. 1
\begin{equation}
\tilde P^{(1)} (\vecc z_\perp; v, \bar v)
=
 - \left(ig \right)^{2} \ (v_\mu \bar v_\nu)  \int_{- \infty}^{\infty}\! d \sigma  \int_{- \infty}^{\infty}\! d \sigma'  \  \mathrm{ Tr} \frac{1}{N_c} \ \langle \star |   \ {\cal P} [ {\cal A}_\mu(v \sigma + \vecc z_{\perp})
 {\cal A}_\nu(\bar v \sigma') ] | \star \rangle \ .
 \label{eq:P_LO}
\end{equation}
\begin{figure}[ht]
 $$\includegraphics[angle=90,width=0.4\textwidth]{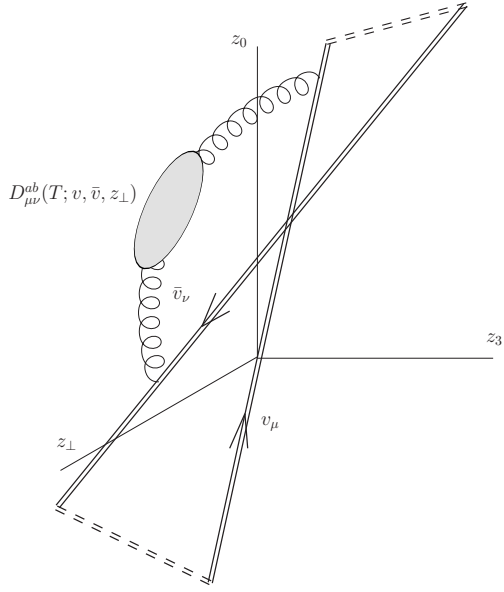}$$
   \caption{Leading-order transverse-distance dependent contribution to the skewed configuration of the Wilson lines in Euclidean space with a generic two-gluon non-thermal small energy scale dependent correlator $D_{\mu\nu}^{ab}(Q_T, z)$. }
\end{figure}

The most general Lorenz and colour structure of the two-gluon non-thermal correlator, which takes into account both perturbative and possible nonperturbative contributions \cite{LN_NP}
\begin{equation}
 \langle \star |   \ [ {A}_\mu^a (v \sigma + \vecc z_{\perp})
 {A}_\nu^b(\bar v \sigma') ] | \star \rangle
=
\delta^{ab} \ D_{\mu\nu} (v \sigma  - \bar v \sigma' + \vecc z_{\perp}) \ ,
\end{equation}
is determined as follows
\begin{eqnarray}
D_{\mu\nu} (z)
 & = &
g_{\mu\nu} \partial^2 D_1(z^2) - \partial_\mu \partial_\nu D_2 (z^2)
 \nonumber \\
& = & g_{\mu\nu} (2 \omega \partial_u + 4z^2 \partial_u^2) D_1(z^2) - (2 g_{\mu\nu} \partial_u + 4 z_\mu z_\nu \partial_u^2) D_2 (z^2) \ ,
\label{eq:2gluon_def}
\end{eqnarray}
where $u \equiv z^2$ and $z = v \sigma + \vecc z_{\perp} - \bar v \sigma'$. Dimension regularisation provides IR finiteness under $\omega = 4 - 2 \varepsilon , \varepsilon < 0$.
We will also make use of the Laplace transform of the functions $D_{1,2}$ and their derivatives $D_{1,2}^{(l)} \equiv \partial_u^l D_{1,2} (u)$ in Euclidean space:
\begin{equation}
D_{1,2}^{(l)}
=
(-1)^l \int_0^\infty \! d\alpha \ \alpha^l {\rm e}^{- \alpha z_\perp^2}\ \tilde D_{1,2} (\alpha) \ .
\end{equation}

Eq. (\ref{eq:P_LO}) can be split up into the following four contributions:
\begin{equation}
\tilde P^{(1)} (\vecc z_\perp; v, \bar v) = g^2 \ C_{\rm F} \ I \ , \  I
 =
  I_{1}+I_{1'}+I_{2}+I_{2'}  \ , \ C_{\rm F} = \frac{N_c^2 - 1}{2N_c} \ ,
\end{equation}
and
\begin{eqnarray}
I_{1} & = & v_{\mu}\bar v_{\nu}\int_{-\infty}^{\infty}d\sigma \int_{-\infty}^{\infty}d\sigma'\ 2\omega g^{\mu\nu}\partial_{u}\, D_{1}(z^{2}) \ , \nonumber \\
I_{1'} & = & v_{\mu}\bar v_{\nu}\int_{-\infty}^{\infty}d\sigma \int_{-\infty}^{\infty}d\sigma'\ 4g^{\mu\nu}z^{2}\partial^{2}_{u}D_{1}(z^{2}) \ ,  \nonumber  \\
I_{2} & = & - v_{\mu}\bar v_{\nu}\int_{-\infty}^{\infty}d\sigma \int_{-\infty}^{\infty}d\sigma'\ 2g^{\mu\nu}\partial_{u}\, D{}_{2}(z^{2}) \ , \nonumber  \\
I_{2'} & = & - v_{\mu}\bar v_{\nu}\int_{-\infty}^{\infty}d\sigma \int_{-\infty}^{\infty}d\sigma' \ 4z^{\mu}z^{\nu}\partial^{2}_{u}D{}_{2}(z^{2}) \  .
\label{eq:line_I}
\end{eqnarray}
Evaluating the line integrals (\ref{eq:line_I}), we use the representation
\begin{equation}
{\rm e}^{-\alpha z^{2}}
=
{\rm e}^{-\alpha L^{2}\left(\sigma+\cos\left(\Delta\phi\right)\sigma'\right)^{2}}{\rm e}^{-\alpha L^{2}\sigma'^{2}\sin^{2}\left(\Delta\phi\right)}{\rm e}^{-\alpha b_{\perp}^{2}}  \ ,
\end{equation}
where $\Delta \phi = \frac{\phi - \bar \phi}{2}$.

After a straightforward, but tedious calculation (see for technical details Ref. \cite{Lau_13}), we finally obtain:
\begin{equation}
\tilde P^{(1)} (\vecc z_\perp; v, \bar v)
 =
g^2 C_{\rm F} 2\pi \frac{\cos \Delta \phi}{|\sin \Delta \phi|}\  \left[(\omega - 2 )D_{1}(\vecc z_{\perp}^{2}; \Lambda)+2\vecc z_{\perp}^{2}D'_{1}(\vecc z_{\perp}^{2}; \Lambda)\right]  \ ,
\label{eq:jqeuclidean}
\end{equation}
where we take into account the IR-cutoff $\Lambda \sim gQ_T$ in $D_1$, which might be IR-singular. In contrast, the UV-finiteness is guaranteed by the effective UV-cutoff $\vecc z_\perp$. Eq. (\ref{eq:jqeuclidean}) is our main result for the skewed probability distribution in Euclidean space-time in the leading non-trivial order. Let us note that the contribution of the second term in the two-gluon correlator, $D_2$, falls out, which is indeed required by gauge invariance.

To establish the connection of this function with the physical jet quenching parameter (\ref{eq:hat_q_def}), it is instructive to rewrite the angular factor in the covariant form
\begin{equation}
 K (v, \bar v)
 \equiv
 \frac{\cos \Delta \phi}{|\sin \Delta \phi|}
=
\frac{(v\cdot \bar v)}{\sqrt{v^2 \bar v^2 - (v\cdot \bar v)^2 }} \ ,
\label{eq:ang_fact}
\end{equation}
which allows us to study in detail the transition to the light-cone Minkowskian layout.

In Minkowski space-time, for each pair of time-like vectors $v$ and $\bar v$ traveling in the opposite time direction, a rest frame can be found in which they are parametrized as follows:
\begin{eqnarray}
v & = L \left(\gamma_{1},-\beta_{1}\gamma_{1}, \vecc 0_\perp\right) & =L \left(\cosh\frac{\psi_{1}}{2},-\sinh\frac{\psi_{1}}{2}, \vecc 0_\perp\right),\label{eq:vwMinkowski}\\
\bar v & =- L \left(\gamma_{2},-\beta_{2}\gamma_{2}, \vecc 0_\perp\right) & =- L \left(\cosh\frac{\psi_{2}}{2},-\sinh\frac{\psi_{2}}{2}, \vecc 0_\perp\right)\ .  \nonumber
\end{eqnarray}
Evaluated in these two vectors, the function $K(v, \bar v)$ reads:
\begin{equation}
K(v,\bar v) =  \frac{v\cdot \bar v}{\sqrt{v^{2}\bar v^{2}-\left(v\cdot \bar v\right)^{2}}}
 =
 -i\frac{\cosh\left(\frac{\psi_{1}-\psi_{2}}{2}\right)}{|\sinh \left(\frac{\psi_{1}-\psi_{2}}{2}\right)|} \ .
\label{eq:KMinkowski}
\end{equation}
In the case we are interested in, i.e. the case $\psi_{1}=\psi_{2}$, this expression is singular. Thus, in the limit of $v$ and $\bar v$ lying
in the opposite space-time direction, the parametrization (\ref{eq:KMinkowski}) of $K(v,\bar v)$ is ill-defined.

Moreover, one has to be careful in using the definition (\ref{eq:ang_fact}) $K(v, \bar v)$ in the light-cone case. To illustrate this, take the parametrization
(\ref{eq:vwMinkowski}) of $v$ and $\bar v$ in the limit of infinite rapidity:
\begin{eqnarray}
v_{\rm LC} & = & \frac{L}{2}\left({\rm e}^{\psi_{1}/2},- {\rm e}^{\psi_{1}/2}, \vecc 0_\perp \right)\ , \label{eq:vwLC}\\
\bar v_{\rm LC} & = & -\frac{L}{2}\left({\rm e}^{\psi_{2}/2}, - {\rm e}^{\psi_{2}/2}, \vecc 0_\perp  \right) \ , \nonumber \\
v_{\rm LC}^{2}& = & \bar v_{\rm LC}^{2}=0 \ . \nonumber
\end{eqnarray}
In the light cone limit $\psi_{1},\psi_{2}\rightarrow\infty$, formula (\ref{eq:KMinkowski}) is clearly ill-defined as well. However,
one cannot simply insert (\ref{eq:vwLC}) into definition (\ref{eq:ang_fact}), since that would yield:
\begin{eqnarray*}
K\left(v\cdot \bar v\right)_{LC} & = & -\frac{L^{2}}{4}\frac{e^{\psi_{1}/2}{\rm e}^{\psi_{2}/2}-{\rm e}^{\psi_{1}/2}{\rm e}^{\psi_{2}/2}}{\sqrt{-\frac{L^{4}}{16}\left({\rm e}^{\psi_{1}/2}{\rm e}^{\psi_{2}/2}-{\rm e}^{\psi_{1}/2}{\rm e}^{\psi_{2}/2}\right)^{2}}}=\frac{0}{0} \ .
\end{eqnarray*}
Thus, although we have a covariant definition (\ref{eq:ang_fact}) of the angular dependence $K(v,\bar v)$, we are facing problems both in
the evaluation of $K(v,\bar v)$ for $v$ and $\bar v$ opposite vectors in Minkowski space-time, and in the evaluation of $K(v,\bar v)$ for vectors
on the light cone.

Let us consider a straightforward solution of this problem, which requires just being more careful when taking the light-cone limit. Indeed,
writing:
\begin{eqnarray*}
v^* & = & \lim_{\epsilon\rightarrow0}\frac{L}{2}\left({\rm e}^{\psi_{1}/2}+\epsilon,- {\rm e}^{\psi_{1}/2}+\epsilon , \vecc 0_\perp \right),\\
\bar v^* & = & \lim_{\delta\rightarrow0}-\frac{L}{2}\left({\rm e}^{\psi_{2}/2}+\delta, - {\rm e}^{\psi_{2}/2}+\delta , \vecc 0_\perp \right),
\end{eqnarray*}
the definition of $K(v,\bar v)$ can be readily used, yielding:
\begin{equation}
K\left(v^*,  \bar v^* \right) =
\lim_{\epsilon,\delta\rightarrow0} \Bigg( -\frac{L^{2}}{4}\frac{\epsilon {\rm e}^{\psi_{2}/2}+\delta {\rm e}^{\psi_{1}/2}}{\sqrt{-\frac{L^{4}}{16}\left(\epsilon {\rm e}^{\psi_{2}/2}+\delta {\rm e}^{\psi_{1}/2}\right)^{2}}} \Bigg) =
i \ .
\label{eq:K_1}
\end{equation}
This method suggests, however, setting the primordial partons off-mass-shell.  An even more straightforward method is to place {\it only one vector on the light cone}, while the other vector remains time-like (that is, we define the {\it skewed} layout), for example:
\begin{eqnarray*}
v & \to & v_{\rm LC} =  \frac{L}{2}\left({\rm e}^{\psi_{1}/2}, - {\rm e}^{\psi_{1}/2},\vecc 0_\perp\right),\\
\bar v & = & - L \left(\cosh\frac{\psi_{2}}{2}, - \sinh\frac{\psi_{2}}{2},\vecc 0_\perp\right).
\end{eqnarray*}
In that case, using the definition of $K(v_{LC},\bar v)$ yields (since
$v_{LC}^{2}=0$):
\begin{eqnarray}
K(v_{LC},\bar v) & = & \frac{v_{\rm LC}\cdot \bar v}{\sqrt{v_{\rm LC}^{2}\bar v^{2}-\left(v_{\rm LC}\cdot \bar v\right)^{2}}} =
i \ .
\label{eq:K_2}
\end{eqnarray}
The last equation suggests that the angular multiplier $K(v, \bar v)$ is invariant in the skewed layout, so that one can readily put the second vector on the light-cone.

To illustrate the consistency of the proposed method, let us evaluate the skewed function $\rho (Q_T; v, \bar v) $ in perturbative vacuum, taking into account the hierarchy of scales. The free gluon propagator is given by
\begin{equation}
 D_1 (z^2)
 =
  -\frac{(- \pi \Lambda^2 z^2)^{\varepsilon}}{16 \pi^2} \ \frac{\Gamma (1 - \varepsilon)}{\varepsilon} \ ,
\end{equation}
where $\Lambda^2$ is an IR scale,
so the perturbative transverse-distance dependent probability distribution (\ref{eq:jqeuclidean}) reads (for comparison, see the LO result for the quark-quark scattering amplitude in Refs. \cite{WL_angular})
\begin{equation}
 \tilde P^{\rm pert.} (\vecc z_\perp^2; v, \bar v)
 =
- C_{\rm F} g^{2} \frac{\cos \Delta \phi}{|\sin  \Delta \phi|} \  \frac{(- \pi \Lambda^2 \vecc z_\perp^2)^{\varepsilon}}{4 \pi} \ \frac{\Gamma (1 - \varepsilon)}{\varepsilon} \ \ .
\label{eq:I_pert}
\end{equation}
Note that $\varepsilon$ is negative.
Eq. (\ref{eq:I_pert}) is IR singular in the limit $\varepsilon \to 0^-$, while the skewed function $\rho_{\rm pert.} (Q_T; v, \bar v) $ is, in contrast, IR-finite but UV-divergent:
\begin{equation}
 \rho_{\rm pert.}^{(1)} (Q_T; v, \bar v)
 =
\frac{1}{L}\ g^2 C_{\rm F}\  \frac{\cos \Delta \phi}{|\sin  \Delta \phi|} \int^{k_\perp^{\rm max.}}\! \frac{d^2 k_\perp}{(2\pi)^2} \ ,
\end{equation}
as one expects in the pure perturbative case. {It is natural to assume that the UV cutoff scales as follows \cite{hat_q_T_line}
$$
k_\perp^{\rm max.} \approx \lambda Q = Q_T \ .
$$
Therefore, taking into account Eqs. (\ref{eq:scales_1}-\ref{eq:scales_3}), one has
\begin{equation}
  \rho_{\rm pert.}^{(1)} (Q_T; v, \bar v)
 =
\frac{g^2 C_{\rm F} }{4 \pi} \  \frac{\cos \Delta \phi}{|\sin  \Delta \phi|} \frac{Q_T^2}{L}
=
\frac{g^2 Q_T C_{\rm F} }{4 \pi} \  \frac{\cos \Delta \phi}{|\sin  \Delta \phi|} Q_T^2 \ .
\label{eq:pert_Q_T}
\end{equation}
}
We have found, therefore, that Eq. (\ref{eq:pert_Q_T}) is consistent with the hierarchy of scales which is adopted in the derivation of the definition (\ref{eq:jetquenchingWilson}) within SCET  \cite{hat_q_T_line}.

Let us now estimate the nonperturbative input to the Euclidean skewed function $\rho (Q_T, v, \bar v)$ making use of the simplest Gaussian two-gluon correlator:
\begin{equation}
D_{1}(z^2)
=
{\rm e}^{- \kappa(g^2, Q_T) z_{\perp}^{2}} \ ,
\end{equation}
where $\kappa(g^2, Q_T)$ {suppresses IR divergences in the large-$z_\perp^2$ domain. Taking into account that the medium is characterized, besides the scale $Q_T$, by the lower scales $m_E \sim g Q_T \gg g_E \sim g^2 Q_T$, and that we work in the energy range $[g Q_T, Q_T]$, one can naturally set the medium IR function to the Debye mass:
\begin{equation}
 \kappa(g^2, Q_T) =
 m_E^2 \sim g^2 Q_T^2
 \label{eq:debye_m}
\end{equation}
Note that in our approach $m_E$ plays exactly the same role as the effective gluon mass in Refs.\cite{hat_q_NP}, that is an IR cutoff. }.
The calculation is straightforward and yields
\begin{equation}
\tilde P^{(1)}(\vecc z_{\perp}; v, \bar v)
=
2\pi g^{2} C_{\rm F} \ \frac{\cos\Delta\phi}{|\sin\Delta\phi|}\left[(\omega-2){\rm e}^{-\kappa(g^2, Q_T) z_{\perp}^{2}}-2\kappa(g^2, Q_T) z_{\perp}^{2}{\rm e}^{-\kappa (g^2, Q_T) z_{\perp}^{2}}\right] \ ,
\label{eq:P_1}
\end{equation}
so that
\begin{eqnarray}
\rho_{\rm NP}^{(1)} (Q_T; v, \bar v)  & = & 
\frac{1}{L}\int\frac{d^{2}k_{\perp}}{\left(2\pi\right)^{2}}k_{\perp}^{2}\int d^{2}z_{\perp} {\rm e}^{ik_{\perp}\cdot z_{\perp}}\left(1+ \tilde P^{(1)}(\vecc z_{\perp}; v, \bar v) \right) \nonumber \\ 
& = & 
\frac{C_{\rm F} g^{2}}{L} 8 \pi \kappa (g^2, Q_T) \omega \frac{\cos\Delta\phi}{|\sin\Delta\phi|}  \  .
\label{eq:rho_1}
\end{eqnarray}
As we know from the previous discussion, expression (\ref{eq:rho_1}), being well-defined in Euclidean space, can be safely transformed to the Minkowski space-time.
Taking into account Eq. (\ref{eq:debye_m}),
we obtain the following result (in four-dimensional Euclidean space $\omega = 4$):
\begin{equation}
 \rho_{\rm NP}^{(1)} (Q_T; v, \bar v)  =
g^{2} Q_T \ {C_{\rm F} } \ 32 \pi m_E^2 \frac{\cos\Delta\phi}{|\sin\Delta\phi|}  \sim g_E^2 m_E^2  .
\label{eq:rho_2}
\end{equation}

It is instructive to compare Eq. (\ref{eq:rho_2}), obtained within non-thermal scale-hierarchical effective theory with some recent results for thermal situations available in the literature. It is known that in the leading order, the jet quenching parameter is given by (see, e.g., Refs. \cite{hat_q_NP, hat_q_LAT})
\begin{equation}
  \hat q
  =
  \int_0^{k_\perp^*}\!
  \frac{d^2 k_\perp}{(2\pi)^2} \ \vecc k_\perp^2 \ {\cal C} (\vecc k_\perp) \ , \ {\cal C} (\vecc k_\perp) = g_E^2 C_{\rm F} \left( \frac{1}{\vecc k_\perp^2} - \frac{1}{\vecc k_\perp^2 + m_E^2} \right) + O(g^4) \ ,
\label{eq:hat_q_NP}
\end{equation}
where the effective coupling and the Debye mass are
\begin{equation}
 g_E^2 = g^2 T + O(g^4T) \ , \ m_E = gT \sqrt{N_c/3 + N_f/6} + O(g^3 T) \ ,
\end{equation}
and $k_\perp^*$ takes care of the UV-divergency.
Accordingly, one has
\begin{equation}
  \hat q (k_\perp^*; g_E, m_E)
  =
  \frac{g_E^2 m_E^2 C_{\rm F}}{2\pi} \ln \frac{|k_\perp^*|}{m_E} \ .
  \label{eq:hat_q_NP_fin}
\end{equation}
{Eq. (\ref{eq:rho_2}) reproduces}, therefore, the scale structure of Eq. (\ref{eq:hat_q_NP_fin}), where
{the energy scale $Q_T \sim L^{-1}$ arises as a ``mock temperature'' of the cold medium having a non-trivial non-perturbative vacuum state.} {On the other hand, the naive perturbative contribution (\ref{eq:pert_Q_T}), although correct in power of $Q_T$, lacks an IR scale and thus fails to match Eq. (\ref{eq:hat_q_NP_fin}). This is not surprising since the pure perturbative treatment is not allowed in the far IR region, where a nonperturbative approach must be adopted. The above comparison demonstrates that an approach based on the Wilson lines techniques within the SCET framework allows one to get correct estimation of the jet quenching (which agrees with one obtained in Refs. \cite{hat_q_NP}) even at zero temperature given that there exists a hierarchy of energy scales $Q \gg Q_T \gg m_E \sim gQ_T \gg g_E \sim g^2 Q_T$ \cite{hat_q_T_line}. This provides us with an explicit proof of the relevance of the SCET approach to the jet quenching study.}
 Let us emphasize that our approach is not related to a specific theory or model of the QCD background, while the use of the effective propagator, Eq. (\ref{eq:2gluon_def}), allows one to take into account the properties of the non-thermal medium in the most general covariant form.

\section{Conclusions and outlook} 
We discussed an approach which allows one to separate the angular dependent factor from the Euclidean Wilson lines correlator in the skewed layout and perform consistently its transition to the light-cone Minkowski space-time. The resulting expression gives us, formally, the jet quenching parameter (\ref{eq:rho_def}) as a function of the two-gluon correlation function {at zero temperature}
$$
\langle \star |   \ [ {A}_\mu^a (z)
 {A}_\nu^b(z') ] | \star \rangle
$$
 expressed in the most general {covariant} form (\ref{eq:2gluon_def}). We paid special attention to the angular structure of the skewed function $\rho^{(1)}(Q_T; v, \bar v)$ and studied its behaviour in vicinity of the ``extremal'' configuration $\bar v = - v \ , \ v^2 = 0$. We demonstrated that an appropriate correspondence between the skewed Euclidean and the light-cone Minkowskian layouts can be established and used in practical calculations. {Note that although we addressed the problem of Euclidean analytical continuation of the light-like correlators from the point of view used mostly to approach the issue of the rapidity singularities in transverse-momentum dependent parton densities, (see, e.g., Refs. \cite{CS_T_line, LC_TMD}), our result (\ref{eq:K_1}, \ref{eq:K_2}), confirms the boost-independence of the analytical continuation observed in \cite{hat_q_NP}. Further analysis is needed, however, in order to ensure the possibility of the Euclidean$\leftrightarrow$Minkowski transition at $T \neq 0$ in our formalism. In this case, the two-point gluon correlation functions at finite temperature must replace the simplest Ansatz (\ref{eq:2gluon_def}), see, e.g., recent preprint \cite{YM_2point}.} 
 
Our formalism can be further applied in lattice simulations and in calculations in the QCD vacuum and nuclear medium models.
{We considered the non-thermal medium, that is $T=0$. Therefore, the presence of low energy scales ${Q_T, g Q_T, g^2 Q_T}$ is essential for the gauge-invariant formulation of the jet quenching process in terms of the light-like Wilson lines. We demonstrated that this definition is consistent with the hierarchy of energy scales suggested by the SCET framework, and yields an appropriate result for a generic (nonperturbative) covariant two-gluon correlation function. For example, the effects of the non-trivial structure of QCD vacuum (e.g., instanton and other nonperturbative models \cite{WL_Eucl}) can be investigated directly by means of the proposed method. The study of the jet quenching in the medium with finite temperature will be reported separately \cite{chlt_finite_T}.}

\vspace{.5cm}
\noindent
{\it Acknowledgements}
We thank T. Mertens and F. Van der Veken for useful stimulating discussions. We are grateful to M. Panero and G. D. Moore for pointing out important works on the subject.

\end{document}